\documentclass[a4paper]{aastex62}

\usepackage{natbib}
\usepackage{xcolor}
\usepackage{hyperref}

\newcommand{\psr}{PSR~J1012$+$5307}
\newcommand{\phsref}{J0958$+$5039}
\newcommand{\frf}{J1118$+$1234}
\newcommand{\pibc}{J101307.3$+$531234}
\newcommand{\ibca}{J101307.4$+$530423}
\newcommand{\ibcb}{J101230.6$+$525826}
\newcommand{\ibcc}{J101204.0$+$531332}
\newcommand{\gw}{gravitational wave}
\newcommand{\rcs}{reduced $\chi^2$}

\newcommand{\maspy}{$\rm mas~yr^{-1}$}
\newcommand{\fsps}{$\rm fs~s^{-1}$}
\newcommand{\msun}{\ifmmode\mbox{M}_{\odot}\else$\mbox{M}_{\odot}$\fi}
\newcommand{\lsun}{\ifmmode\mbox{L}_{\odot}\else$\mbox{L}_{\odot}$\fi}
\newcommand{\rsun}{\ifmmode\mbox{R}_{\odot}\else$\mbox{R}_{\odot}$\fi}
\newcommand{\psrpi}{\ensuremath{\mathrm{PSR}\pi}}
\newcommand{\mspsrpi}{\ensuremath{\mathrm{MSPSR}\pi}}

\graphicspath{{./}{figures/}}

\accepted{April 29, 2020}

\submitjournal{ApJ}

\shorttitle{VLBI astrometry of \psr}
\shortauthors{Ding et~al.}

\begin{document}

\title{Very long baseline astrometry of \psr\ and its implications on alternative theories of gravity}

\email{haoding@swin.edu.au}

\author{Hao Ding}
\affil{Centre for Astrophysics and Supercomputing, Swinburne University of Technology \\
John St, Hawthorn, VIC 3122, Australia}
\affil{ARC Centre of Excellence for Gravitational Wave Discovery (OzGrav)}

\author{Adam T. Deller}
\affil{Centre for Astrophysics and Supercomputing, Swinburne University of Technology \\
John St, Hawthorn, VIC 3122, Australia}
\affil{ARC Centre of Excellence for Gravitational Wave Discovery (OzGrav)}

\author{Paulo Freire}
\affil{Max-Planck-Institut f$\ddot u$r Radioastronomie, Auf dem H$\ddot u$gel 69, D-53121 Bonn, Germany}

\author[0000-0001-6295-2881]{David L. Kaplan}
\affil{Department of Physics, University of Wisconsin, Milwaukee, WI 53201, USA}

\author{T. Joseph W. Lazio}
\affil{Jet Propulsion Laboratory, California Institute of Technology, 4800 Oak Grove Dr, Pasadena, CA 91109, USA}

\author{Ryan Shannon}
\affil{Centre for Astrophysics and Supercomputing, Swinburne University of Technology \\
John St, Hawthorn, VIC 3122, Australia}\}
\affil{ARC Centre of Excellence for Gravitational Wave Discovery (OzGrav)}
\author{Benjamin Stappers}
\affil{University of Manchester, Jodrell Bank Centre for Astrophysics, Alan Turing Building, Manchester M13 9PL}




\begin{abstract}

\psr, a millisecond pulsar in orbit with a helium white dwarf (WD), has been timed with high precision for about 25 years. One of the main objectives of this long-term timing is to use the large asymmetry in gravitational binding energy between the neutron star and the WD to test gravitational theories.
Such tests, however, will be eventually limited by the accuracy of the distance to the pulsar. 
Here, we present VLBI (very long baseline interferometry) astrometry results spanning approximately 2.5 years for \psr, obtained with the Very Long Baseline Array as part of the \mspsrpi\ project. These provide the first proper motion and absolute position for \psr\ measured in a quasi-inertial reference frame.
From the VLBI results, we measure a distance of $0.83^{+0.06}_{-0.02}$\,kpc (all the estimates presented in the abstract are at 68\% confidence) for \psr, which is the most precise obtained to date. Using the new distance, we improve the uncertainty of measurements of the unmodeled contributions to orbital period decay, which, combined with three other pulsars, places new constraints on the coupling constant for dipole gravitational radiation $\kappa_D=(-1.7\pm1.7)\times 10^{-4}$ and the fractional time derivative of Newton's gravitational constant $\dot{G}/G = -1.8^{\,+5.6}_{\,-4.7}\times 10^{-13}\,{\rm yr^{-1}}$ in the local universe. 
As the uncertainties of the observed decays of orbital period for the four leading pulsar-WD systems become negligible in $\approx10$ years, the uncertainties for $\dot{G}/G$ and $\kappa_D$ will be improved to $\leq1.5\times10^{-13}\,{\rm yr^{-1}}$ and $\leq1.0\times10^{-4}$, respectively, predominantly limited by the distance uncertainties.

\end{abstract}


\section{Introduction} \label{sec:intro}

\subsection{Testing gravity theories with millisecond pulsars} \label{subsec:intro_test_gravity_with_psr}
When the beams of radiation emitted by rotating neutron stars (NSs) sweep across our line of sight, the result is a regular, lighthouse-like train of pulses. 
Thanks to their high rotational inertia, the spin period is extremely stable. The difference between observed pulse times of arrival (ToAs) and the model prediction for those ToAs is known as their residuals. The residuals are generally used to study unmodeled or imperfectly modeled physical process that would affect the ToAs \citep[e.g.][]{Lorimer12}.
Such processes include propagation through ionised material surrounding a binary companion \citep[e.g.][]{Lyutikov05} or through the interstellar medium \citep[ISM; e.g.][]{Lyne68,Bhat04} and gravitational phenomena.

Einstein's theory of general relativity (GR) is the simplest possible form among a class of candidate gravitational theories. Some alternatives to general relativity suggest a time dependence of Newton's gravitational constant $G$, which in most cases also necessitates dipolar \gw\ emission \citep{Will93}. There are several ways to test GR and constrain alternative theories of gravity with pulsars. 
Highly relativistic double neutron star systems \citep[e.g.][]{Damour92,Burgay03}, pulsar-white dwarf binaries \citep[e.g.][]{Lazaridis09,Freire12,Antoniadis13,Zhu15} and triple stellar systems hosting pulsars \citep{Archibald18} have probed different regions of phase space for deviations from the predictions of general relativity.
Taken collectively, an ensemble of pulsars can be used as a Pulsar Timing Array (PTA) to search for spatially correlated ToA variations that would betray the presence of nanohertz gravitational wave, such as those generated by supermassive binary black holes \citep{Detweiler79}.

The pulsars used in these tests belong to a sub-group of pulsars called “recycled” or ``millisecond” pulsars (MSPs).
In this work, we use these two terms interchangeably to refer to pulsars spun up via accretion from a companion donor star \citep{Alpar82}. 
One of the reasons why MSPs are important for these experiments is that they exhibit less intrinsic timing noise compared to non-recycled pulsars \citep{Shannon10}, thus providing much better timing stability \citep[e.g.][]{Perera19} and higher timing precision; this is important for the detection of small relativistic effects in their orbits.

In addition to studies based on pulsar timing, Very Long Baseline Interferometry (VLBI) astrometric experiments are also carried out on both MSPs and normal pulsars \citep[e.g.][]{Chatterjee09,Deller16,Vigeland18,Deller19}. 
By measuring annual geometric parallax and proper motion, VLBI astrometry can achieve model-independent estimates of distance, transverse velocity, and absolute positions for pulsars. It is significant not only in reducing the distance uncertainty and improving sensitivities of PTAs \citep{Madison13}, but also in various applications on a case by case basis.

\subsection{The \psr\ binary system}
\label{subsec:intro_J1012}
\psr\ is a millisecond pulsar (MSP) with a 5.3\,ms rotational period \citep{Nicastro95}. It has a helium white dwarf (WD) companion with mass $0.156\pm0.020$\,\msun\ (at 68\% confidence level, as is any other quoted uncertainty in this paper unless otherwise stated, \citealp{van-Kerkwijk96,Callanan98}) in a 0.6-day-long \citep{Lazaridis09} near-circular ($e<8\times10^{-7}$) orbit at a moderate inclination angle \citep{Driebe98,Lange01}. 
Spectroscopic observations of the WD were used to measure the mass ratio between \psr\ and the WD companion to be $10.5\pm0.5$ by \citet{Callanan98} and $10.0\pm0.7$ by \citet{van-Kerkwijk04}. 
A new spectroscopic study of \psr\ by \citet{Mata-Sanchez20} further refines the mass ratio to $10.44\pm0.11$.
\citet{Antoniadis16} re-visited the \citet{Callanan98} data with an updated model for the helium WD, which resulted in an updated mass estimate of $0.174\pm0.011$\,\msun. 
In this paper, we will use the new mass ratio $10.44\pm0.11$ and WD mass $0.174\pm0.011$\,\msun, that corresponds to a mass estimate of $\sim$1.8\,\msun\ for the NS in \psr.

\subsection{Motivations for improving the distance to \psr}
\label{subsec:intro_motivation}
The large difference in gravitational binding energy between the NS and WD in the \psr\ system means that it would be an efficient emitter of dipolar gravitational waves in some alternate theories of gravity.
Timing observations of such binary pulsars are able to offer experimental tests for those theories.
Incorporating timing data from both \psr\ and PSR J0437$-$4715, \citet{Lazaridis09} looked into the contributions to the time derivative of orbital period $\dot P_b$ and use the difference between the modeled and observed value $\dot P_b^{\,\mathrm{obs}}$ to constrain the coupling constant for  dipole gravitational radiation $\kappa_D$ to be $(0.3\pm2.5)\times10^{-3}$ and the fractional time derivative of Newton's gravitational constant $\dot{G}/G$ to be $(-0.7\pm3.3)\times10^{-12}~{\rm yr}^{-1}$, both at 95\% confidence. 
Using the same method but different pulsar-WD binaries, the best pulsar-based constraints, $\kappa_D=(-0.3\pm2.0)\times10^{-4}$ (at 68\% confidence) and $\dot{G}/G=(-1\pm9)\times10^{-13}\,{\rm yr^{-1}}$ (at 95\% confidence) are derived, respectively, by \citet{Freire12} and \citet{Zhu18}.

For \psr\ the precision of the constraints on $\kappa_D$ and $\dot G$ is dominated by the uncertainty in $\dot P_b^{\,\mathrm{obs}}$,
and the distance to \psr\ \citep{Lazaridis09}. 
Accordingly, improvements in the precision of distance estimates to pulsars such as \psr\ have great potential to improve tests of alternate gravitational theories.
Furthermore, as the uncertainty of $\dot P_b^{\,\mathrm{obs}}$ decreases much faster than the uncertainty on the distance with pulsar-timing observations \citep{Bell96}, the latter will eventually dominate the error budget of $\kappa_D$ and $\dot G$.

Improving the distance to \psr\ will also benefit some other studies. 
Additionally, as one of the pulsars having reliable independent distance measurements, \psr\ was used by \citet{Yao17} to derive the latest model of the Galactic free electron density distribution. Therefore a more accurate distance to \psr\ would further refine such a model.

\subsection{Measuring the distance to \psr}
\label{subsec: distance2psr}
Several methods have been used in the past to estimate the distance to \psr, the results of which are summarized in Table~\ref{tab:astrometric_fits}.
A measure of $0.84\pm0.09$\,kpc was derived by \citet{Callanan98} using optical spectral-line observations of the WD companion. Pulsar timing is another way to measure the distance, as timing parallax is one of the outputs from parameter fits. \psr\ is routinely timed by two PTAs, the European Pulsar Timing Array (EPTA) and the North American Nanohertz Observatory for Gravitational Waves (NANOGrav). To date, two timing parallaxes have been reported for \psr\ utilizing solely EPTA data. 
\citet{Lazaridis09} reported a timing parallax of $1.22\pm0.26$\,mas using 15 years of multi-telescope data, corresponding to a distance of $0.82^{+0.22}_{-0.14}$\,kpc. A different timing parallax for \psr\ $0.71\pm0.17$\,mas is reached more recently in \citet{Desvignes16}, showing $1.6\,\sigma$ tension with the previous timing result.
The EPTA and NANOGrav data for \psr\ are also combined and analysed collectively under the International Pulsar Timing Array (IPTA) collaboration, leading to a distance estimate $0.7^{\,+0.2}_{\,-0.1}$\,kpc for \psr\ \citep{Verbiest16}.

Apart from the above-mentioned methods, high-resolution trigonometric astrometry with VLBI or optical observations are able to provide model-independent distance estimation. A reliable distance to \psr\ is essential for improving the uncertainty of $\dot G$ and dipole \gw\ emission (see Section~\ref{subsec:alternative_gravity} for explanation). 
As well as GR tests, a distance based on trigonometric parallax reinforces the estimation of the bolometric luminosity of the companion WD, which reveals the WD radius. The WD radius can be translated to the WD mass when the mass-radius relation is worked out. Incorporating the known NS-WD mass ratio \citep{Mata-Sanchez20}, we can estimate the NS mass.

Prior to this work, an optical counterpart for \psr\ has been identified by \citet{Jennings18} in the Gaia second data release (DR2) \citep{Gaia-Collaboration16,Gaia-Collaboration18}, carrying a tentative parallax of $1.3\pm0.4$\,mas. The parallax was then translated into a Gaia distance $0.79^{+0.73}_{-0.09}$\,kpc for \psr\ incorporating other prior information \citep{Jennings18}.
In this work we focus on VLBI astrometry of \psr\ as part of the \mspsrpi\ project, which is the extension of \psrpi\ project \citep{Deller11,Deller16,Vigeland18,Deller19} focusing exclusively on MSPs. 
Throughout this paper, parameter uncertainties are quoted to 68\% confidence level unless stated otherwise.

\section{Observations and data reduction} \label{sec:observation}
\psr\ was observed at L band (central frequency $\sim$1550 MHz) with the Very Long Baseline Array (VLBA) in eight epochs between July 2015 and November 2017 under the observation codes BD179 and BD192.  Each session was 1\,h long, and the observations are summarized in Table~\ref{tab:eight_positions}. The observational setup is in general the same as other pulsars in \mspsrpi\ and \psrpi\ sample \citep[refer to][]{Deller19}, while using \phsref\ as phase reference calibrator and \frf\ to calibrate the instrumental bandpass. Four compact extragalactic radio sources within 9 arcminutes of \psr\ were identified as suitable in-beam calibrators in early \mspsrpi\ observations, from which \pibc\ is chosen as the primary in-beam calibrator due to its relative brightness (Table~\ref{tab:source_catalog}). 
The data were correlated using the DiFX software correlator \citep{Deller11a}
in two passes - gated and ungated. After gating, the S/N increases by $\approx40$\%.
The gated visibility datasets are subsequently processed in a python-based ParselTongue \citep{Kettenis06} pipeline calling {\tt AIPS} \citep{Greisen03} and {\tt DIFMAP} \citep{Shepherd94} functions, described in \citet{Deller19}. 
The reduction pipeline is publicly available now at \url{https://github.com/dingswin/psrvlbireduce}. It is to be released incorporating better readability, configurability, some new functions and extended diagnostic tools. The data for \psr\ were reduced using the pipeline versioned a6b666e.
Multiple runs of the pipeline are made to iteratively flag bad visibility data (e.g., due to radio frequency interference), make uniform models for calibrators (including the phase calibrator, fringe finder and in-beam calibrators) and obtain reliable positions for \psr.

\begin{table*}
	\centering
	\caption{Positions and uncertainties without $\mid$ with systematics}
	\label{tab:eight_positions}
	\begin{tabular}{ccccc} 
		\hline
		\hline
		yyyy-mm-dd & Project & S/N & RA & Dec \\
		 & Code & & (J2000) & (J2000)\\ 
		\hline
		2015-07-16 & bd179e0 & 58.9 & $10^{\rm h}12^{\rm m}33\fs 439455(5|8)$ & 53\degr07'02\farcs14401(9$\mid$16)\\ 
		2015-11-15 & bd179e1 & 96.8 & $10^{\rm h}12^{\rm m}33\fs 439723(2|6)$ & 53\degr07'02\farcs13524(6$\mid$16) \\
		2016-11-11 & bd192e0 & 87.7 & $10^{\rm h}12^{\rm m}33\fs 440038(3|6)$ & 53\degr07'02\farcs11042(6$\mid$16) \\
		2016-11-19 & bd192e1 & 85.0 & $10^{\rm h}12^{\rm m}33\fs 440040(3|9)$ & 53\degr07'02\farcs10984(8$\mid$20) \\
		2017-05-10 & bd192e2 & 88.2 & $10^{\rm h}12^{\rm m}33\fs 439927(3|7)$ & 53\degr07'02\farcs09865(7$\mid$19) \\
		2017-05-29 & bd192e3 & 46.8 & $10^{\rm h}12^{\rm m}33\fs 439949(5|7)$ & 53\degr07'02\farcs09682(12$\mid$19) \\
		2017-06-11 & bd192e4 & 31.0 & $10^{\rm h}12^{\rm m}33\fs 439972(8|10)$ & 53\degr07'02\farcs09639(19$\mid$23) \\
		2017-11-26 & bd192e5 & 17.9 & $10^{\rm h}12^{\rm m}33\fs 440345(13|15)$ & 53\degr07'02\farcs08382(34$\mid$38)  \\
		\hline
	\end{tabular}
\end{table*}

\begin{table*}
	\centering
	\caption{Source catalog}
	\label{tab:source_catalog}
	\begin{tabular}{cccc} 
		\hline
		\hline
		Source & Name in data & $\Delta_{\rm src-psr}$ & Purpose\\
		\hline
		\psr\ & J1012$+$5307 & 0 & target \\
		\frf\ & \frf\ & 42\fdg6 & fringe finder/bandpass calibrator\\ 
		\phsref\ & \phsref\ & 3\fdg26 & phase calibrator \\
		\pibc\ & IBC00462 & 7\farcm51 & primary in-beam calibrator\\
		\ibca\ & IBC00412 & 5\farcm74 & in-beam calibrator\\
		\ibcb\ & IBC00421 & 8\farcm62 & in-beam calibrator\\
		\ibcc\ & IBC00460 & 7\farcm85 & in-beam calibrator\\
		\hline
	\end{tabular}
\end{table*}

\section{Systematic errors and parallax fits} \label{sec:result}
After data reduction, we determined positions of \psr\ at eight epochs, which are summarized in Table~\ref{tab:eight_positions}. The statistical positional uncertainties obtained from an image-plane fit are reported to the left of the "$\mid$" symbol in Table~\ref{tab:eight_positions}.  However, we expect a significant contribution from systematic position shifts, and the uncertainties in Table~\ref{tab:eight_positions} to the right of "$\mid$" symbol incorporate both the statistical uncertainty already mentioned and an empirical estimate of systematic uncertainty.  This estimate is made using the following empirical function rewritten from \citet{Deller19}:
\begin{equation}
\label{eq:empirical_sys_error}
\Delta_{sys}=A \times s \times \overline{\csc{\theta}} + B/S\,,
\end{equation}
where $\Delta_{sys}$ is the ratio of the systematic error to the synthesized beam size, $\theta$ stands for elevation angle, $\overline{\csc{\theta}}$ is the average $\csc{\theta}$ for a given observation (over time and antennas), $s$ is the angular separation in arcmin between \pibc\ and \psr, $S$ represents the signal-to-noise ratio of \pibc, and $A=0.001$ and $B=0.6$ are empirically determined coefficients based on the \psrpi\ sample. The first term in Equation~\ref{eq:empirical_sys_error} represents propagation-related systematic errors, while the second term accounts for random errors resulting from the calibration solutions from the primary in-beam calibrator. In general, propagation-related systematic errors (proportional to the separation between calibrator and target) dominate the systematic (and indeed overall) error budget. This is still true in this work, even if the first term is significantly reduced by the usage of in-beam calibrators. 

For \psr, the contribution from the first term is $\sim$3 times that of the second term. 
The full uncertainties are the addition in quadrature of statistical and systematic uncertainties. The inclusion of systematic uncertainties decreases the \rcs\ of least-squares astrometric fit from 7.9 to 1.9, which indicates that the uncertainty estimation is likely reasonable (although perhaps still modestly underestimated).

\subsection{Astrometric fitting} \label{subsec:pi&pm}
\begin{figure*}
\centering
\includegraphics[width=12cm]{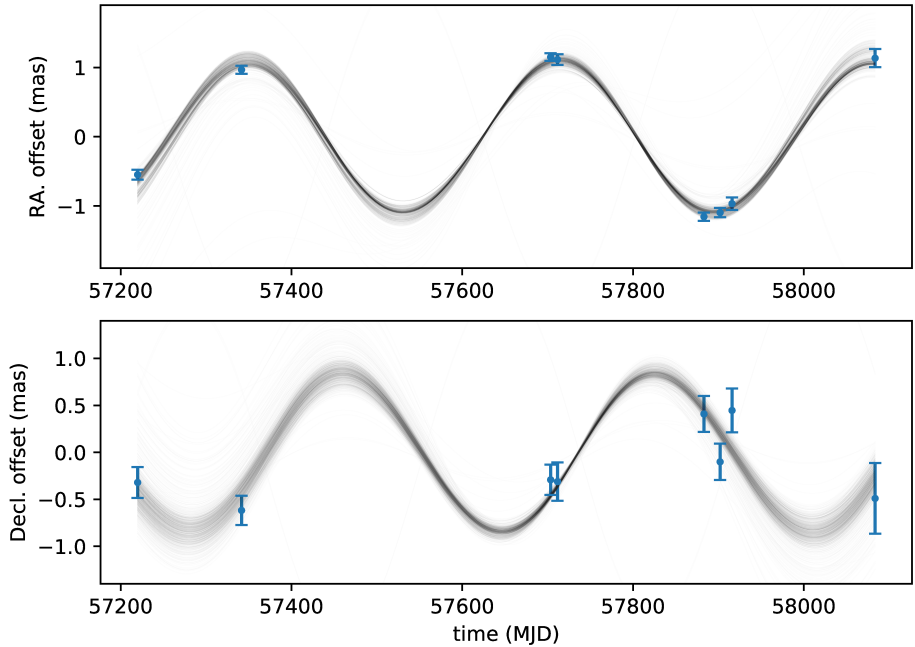}
\caption{Stacked sky position evolution removing best fit proper motion. Each line shows the fitted model out of a bootstrap run.}
\label{fig:bootstrap_parallax}
\end{figure*}

\begin{figure*}
\centering
\includegraphics[width=18cm]{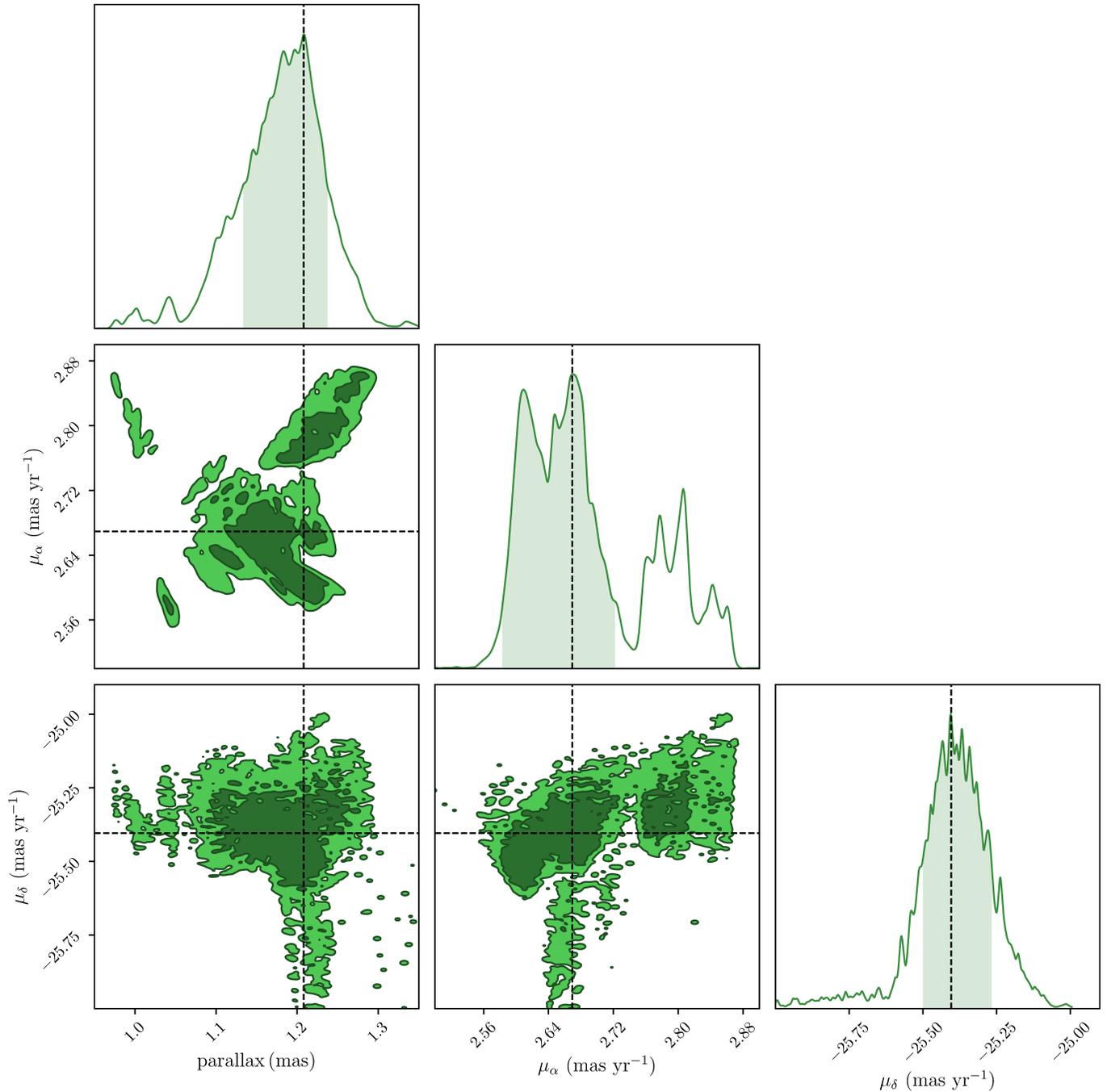}
\caption{
Error ``ellipses'' and marginalized histograms for parallax and proper motion.
In each histogram, the dashed line marks the measured value; the shade stands for the 68\% confidence interval.
In each error ``ellipse'', the dark and bright contour(s) enclose, respectively, 68\% and 95\% of the bootstrapped data points.
}
\label{fig:pdf_pyramid}
\end{figure*}

We used {\tt pmpar}\footnote{\url{https://github.com/walterfb/pmpar}} to perform astrometric fitting for parallax, proper motion and reference position. As found by \citet{Deller19} for PSR~B1913$+$16, using a bootstrap technique to estimate the astrometric parameters was consistent with but more conservative than a least-squares fit, and we followed this approach for \psr. We bootstrapped 130000 times from the 8 positions with full uncertainties (i.e., in every run we drew positions eight times with replacement from the set of positions and performed astrometric fitting on the drawn sample) for estimation of the astrometric parameters. It is possible that the eight draws contain fewer than three effective epochs, which is the minimum required for astrometric fitting. In order to limit the number of severely biased fits yielded by overly short time baselines or negligible variation in the contribution of parallax between epochs,
we required at least 4 effective epochs for subsequent fitting in every bootstrap run.
Figure~\ref{fig:bootstrap_parallax} shows the stacked position evolution removing best fitted proper motion. Each line is the fitting result out of one bootstrap run. The densest part corresponds to the estimated parallax, position and proper motion. 
Using the probability density functions of parallax and proper motion obtained from the bootstrap runs (see Figure~\ref{fig:pdf_pyramid}), we determined the measured value and corresponding uncertainty, which are summarized in Table~\ref{tab:astrometric_fits} along with the results of a simple linear least squares fit. 
For each astrometric observable (i.e. parallax, proper motion, or reference position), its measured value corresponds to the peak of its probability density function; its uncertainty is given by the narrowest interval that encloses 68.3\% of its 130000 bootstrapped results. 

As is seen in Table~\ref{tab:astrometric_fits}, the uncertainties from bootstrap are much more conservative than direct fitting, and are possibly over-estimated since the \rcs\ for direct fitting is already close to unity. We use the bootstrap results in the following discussions. The histogram for $\mu_\alpha$ is bimodal in Figure~\ref{fig:pdf_pyramid}, which is discussed in Section~\ref{subsec:timing_interplay}. Lutz-Kelker correction was not carried out since our significance of parallax $\varpi_0/\sigma>16$ is well over the critical value between 5.0 and 6.7, indicating a negligible Lutz-Kelker effect \citep{Lutz73}. 

\begin{deluxetable*}{ccccccc}[b!]
\tablecaption{Parallax, proper motion, parallactic distance and transverse velocity for \psr\ 
\label{tab:astrometric_fits}}
\tablehead{
\colhead{method} & \colhead{$\varpi$} & \colhead{$\mu_\alpha \equiv \dot{\alpha} \cos\delta$} & \colhead{$\mu_{\delta}$} & \colhead{$D$} & \colhead{$v_t$} & \colhead{References \tablenotemark{a}}\\
& \colhead{(mas)} & \colhead{($\rm mas~{yr}^{-1}$)} & \colhead{($\rm mas~{yr}^{-1}$)} & \colhead{(kpc)} & \colhead{($\rm km~s^{-1}$)} &
}
\startdata
direct fitting & $1.17\pm 0.02$ & $2.68\pm 0.03$ & $-25.38\pm 0.06$ & $0.86\pm 0.02$ & $103.4\pm 1.9$ & This work\\
bootstrap & $1.21^{+0.03}_{-0.08}$ & $2.67^{+0.05}_{-0.09}$ & $-25.40^{+0.14}_{-0.09}$ & $0.83^{+0.06}_{-0.02}$ & $100.2^{+7.2}_{-2.7}$ & This work\\
\\
timing parallax (EPTA/2009) & \textcolor{black}{$1.22\pm 0.26$} & \textcolor{black}{$2.56\pm 0.01$} & \textcolor{black}{$-25.61\pm 0.02$} & $0.8\pm 0.2$ & $100.0\pm 21.3$ \tablenotemark{b} & (1)\\
timing parallax (EPTA/2016) & \textcolor{black}{$0.71\pm 0.17$} & \textcolor{black}{$2.61\pm 0.01$} & \textcolor{black}{$-25.48\pm 0.01$} & $1.4^{+0.4}_{-0.3}$ \tablenotemark{c} & $171.0\pm 41.0$ & (2) \\
orbital parallax (EPTA/2016) & $\cdots$ & $\cdots$ & $\cdots$ & $0.94\pm 0.03$ \tablenotemark{d} & $\cdots$ & (2) \\
timing (IPTA) & $-$ & $-$ & $-$ & $0.7^{+0.2}_{-0.1}$ & $-$ & (3) \\
timing (NANOGrav) & $1.3\pm 0.4$ \tablenotemark{e} & $2.66\pm 0.03$ & $-25.50\pm 0.04$ & $>0.5$ & $-$ & (4) \\
Gaia DR2 & $1.33\pm 0.41$ & $2.98\pm 0.52$ & $-26.94\pm 0.63$ & $0.79^{+0.73}_{-0.09}$ & $113^{+133}_{-12}$ & (5,~6,~7)\\
spectroscopy & $-$ & $-$ & $-$ & $0.84 \pm 0.09$ & $-$ & (8)\\
\enddata
\tablenotetext{a}{ (1)~\citet{Lazaridis09}, (2)~\citet{Desvignes16}, (3)~\citet{Verbiest16}, (4)~\citet{Arzoumanian18}, (5)~\citet{Gaia-Collaboration16}, (6)~\citet{Gaia-Collaboration18}, (7)~\citet{Jennings18}, (8)~\citet{Callanan98}.}
\tablenotetext{b}{ Here, we have re-calculated $v_t$ using the parallax-based distance, rather than weighted distance for consistency \citep{Lazaridis09}.}
\tablenotetext{c}{ $D_\varpi$: Lutz-Kelker correction not applied for consistency.}
\tablenotetext{d}{ $D_{\dot P_b}$, distance derived from the $\dot P_b^{\,\mathrm{obs}}$ (time derivative of orbital period) budget in GR regime.}
\tablenotetext{e}{ Classified by \citet{Arzoumanian18} as non-detection.}
\end{deluxetable*}

\subsection{Absolute position for \psr} \label{subsec:abspos}
\begin{figure}
\centering
\includegraphics[width=14cm]{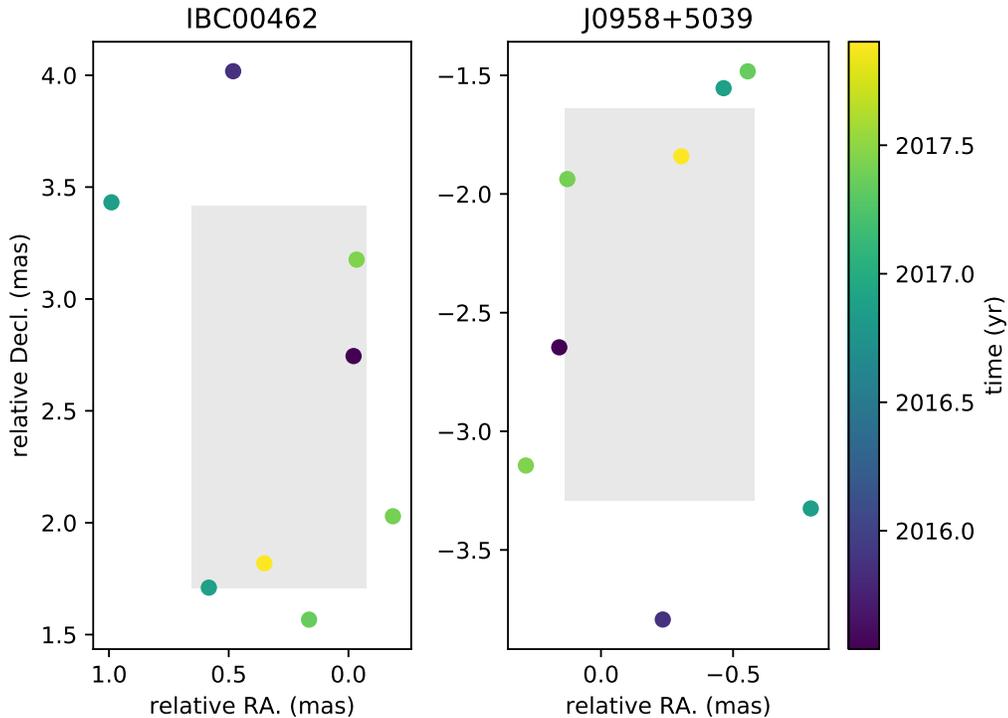}
\caption{Position scatter of the phase calibrator in reference to the primary in-beam calibrator (left) and primary in-beam calibrator referenced to the phase calibrator (right). The positions are relative to $09^{\rm h}58^{\rm m}37\fs 80944+$50\degr39'57\farcs4837 for J0958$+$5039 and $10^{\rm h}13^{\rm m}07\fs29548+$53\degr12'34\farcs3348 for IBC00462. The shaded rectangle in each panel shows the standard deviation of the position in RA and Dec.
\label{fig:position_scatter}}
\end{figure}

Absolute positions are of significance for comparing positions based on different reference frames.
The position for \psr\ that we obtain from bootstrapping is anchored to \pibc, the primary in-beam calibrator. However, the absolute position of \pibc\ is not well determined, and we must estimate its position and uncertainty based on the \mspsrpi\ observations. In order to derive the absolute position for \psr, we used \pibc\ to tie \psr\ to \phsref. We used two methods to make this connection. In the first approach the calibration solutions derived by \phsref\ were transferred to \pibc, which was subsequently divided by the uniform \pibc\ model obtained from eight epochs; the centroid of the divided \pibc\ was located for each epoch; the average and scatter of the eight \pibc\ positions thus offer the information of the ``real" \pibc\ position relative to \phsref\ and the systematic uncertainty of this position. The second way is in principle the same while in the reverse direction: the final solution derived by \pibc\ was applied to \phsref. Figure~\ref{fig:position_scatter} shows the eight \pibc\ and \phsref\ positions obtained in the two different ways. As is expected, from Figure~\ref{fig:position_scatter} we can conclude that 1) no time-dependence of position shifts is noticeable; 2) the scatter among the positions for the two objects, indicating the systematic errors around the mean position, is consistent in both RA and Dec. 
In each way we used the average position to tie \phsref\ and \pibc, thus anchoring \psr\ to \phsref. The absolute positions derived from two ways are highly consistent; we proceed with their average position. Finally, we aligned \psr\ to the latest \phsref\ position\footnote{\url{http://astrogeo.org/vlbi/solutions/rfc_2019a/rfc_2019a_cat.html}}, which is measured at higher radio frequencies based primarily on dual-band 2.3/8.4 GHz observations \citep{Petrov08}. 

As the jet core of \phsref\ is presumably the brightest spot in the \phsref\ map, it is taken as the reference position for \phsref\ after fringe fitting in {\tt AIPS}. Since the jet core moves upstream towards the central engine with increasing frequency \citep[e.g.][]{Bartel86,Lobanov98}, our presented absolute position is referenced to the jet core of \phsref\ at L band, where its absolute position has not been determined.
\citet{Sokolovsky11} compiled multi-band observations on 20 AGNs and reported the median core shift between X band and L band is 1.15\,mas. A recent work by \citet{Plavin19} integrates long-term observations of 40 AGNs and concludes the core shift of AGNs between 8\,GHz and 2\,GHz is typically 0.5\,mas. They additionally found time variability of core shift at an average level of 0.3\,mas in 33 AGNs of the sample. With limited knowledge about the core shift of \phsref, 
we split the median core shift 1.15\,mas evenly between the two axes and add them in quadrature to the errors of the absolute position of \psr.

The absolute position we obtained for \psr\ is shown in Table~\ref{tab:absolute_position}.
We chose the midpoint of the eight VLBI epochs as the reference time for astrometric fitting to obtain the highest precision for the absolute position of \psr, and note that extrapolating the position to earlier or later times will suffer progressively from the accumulation of proper motion uncertainty.

The uncertainty of the absolute position of \psr\ comprises the bootstrap uncertainty of \phsref\ position anchored to \pibc\ (i.e. the uncertainty derived from the normalized histogram of RAs/Decs as shown in Figure~\ref{fig:pdf_pyramid}), the systematic errors in \phsref$-$\pibc\ connection (the scatter of eight positions) and the uncertainty of the absolute position of \phsref. These components are added in quadrature. In order to make comparison to timing results, we also extrapolated the timing positions for \psr\ to our reference epoch MJD~57700 using the ephemerides of \citet{Lazaridis09,Desvignes16,Arzoumanian18} (see Table~\ref{tab:absolute_position}). Furthermore, we re-identified Gaia DR2 851610861391010944 as the optical counterpart for \psr: its predicted position at MJD~57700 is $<5$\,mas from our VLBI position (as shown in Table~\ref{tab:absolute_position}), while its proper motion and parallax are largely consistent with both the VLBI and timing results (Table~\ref{tab:astrometric_fits}). 
The uncertainties of the timing and Gaia positions are estimated with Monte-Carlo simulation, assuming the astrometric parameters offered in literature (also reproduced in Table~\ref{tab:astrometric_fits}) follow a gaussian distribution.

\begin{table*}
	\centering
	\caption{Absolute position for \psr\ at MJD\,57700}
	\label{tab:absolute_position}
	\begin{tabular}{cccccc} 
		\hline
		\hline
		& This work & \citet{Lazaridis09} & \citet{Desvignes16} & \citet{Arzoumanian18} & Gaia DR2 \\
		\hline
	RA & $10^{\rm h}12^{\rm m}33\fs 
	4399(1)$ \tablenotemark{a}
	& $10^{\rm h}12^{\rm m}33\fs 43967(4)$ & $10^{\rm h}12^{\rm m}33\fs 43973(2)$ & $10^{\rm h}12^{\rm m}33\fs 439773(9)$ & $10^{\rm h}12^{\rm m}33\fs 43986(9)$ \tablenotemark{b} \\
	Dec. & 
	53\degr07'02\farcs113(1) &
	53\degr07'02\farcs1094(4) & 53\degr07'02\farcs1113(1) & 53\degr07'02\farcs11090(9) & 53\degr07'02\farcs1098(9) \\
	\hline
	\end{tabular}
	\tablenotetext{a}{The uncertainty for both R.A. or declination includes an estimate of the systematic error introduced by core shift in the reference source between 1.5 GHz and 8.4 GHz, taken as 0.8\,mas in each axis as described in the text.}
	\tablenotetext{b}{\citet{Gaia-Collaboration16,Gaia-Collaboration18,Jennings18}}
\end{table*}

\section{Discussion} \label{sec:discussion}
\subsection{Comparison to timing astrometry} \label{subsec:timing_interplay}
There are two published parallaxes and three proper motions for \psr\ based on timing astrometry \citep{Lazaridis09,Desvignes16,Arzoumanian18}. The timing proper motions disagree significantly, as shown in Table~\ref{tab:astrometric_fits}, indicating that the uncertainties have historically been somewhat underestimated. 
At the time of writing, the EPTA is the only PTA that detects a timing parallax for \psr\ \citep{Lazaridis09,Desvignes16}.
Given the additional data available to the 2016 work, we would expect this to be the more accurate of the two EPTA results.
Our proper motion and reference position agree with both EPTA measurements, 
slightly favoring the 2016 measurement. 
Our measured parallax, on the other hand, agrees with the 2009 measurement but is in significant tension ($\approx2.6\,\sigma$) to the 2016 result. Table~\ref{tab:astrometric_fits} shows that the \citet{Desvignes16} result is also inconsistent with other independently derived distance measurements, but the cause of the discrepancy is unknown.

When we consider the VLBI results for proper motion, a bimodality is
apparent in the probability density obtained for $\mu_\alpha$ in Figure~\ref{fig:pdf_pyramid}.
The sub-peak of $\mu_\alpha$ at $\approx$2.8\,\maspy\ is strongly disfavored by all timing results (cf. Table~\ref{tab:astrometric_fits}). 
We diagnosed the origin of this bimodality by removing one epoch at a time from our bootstrap procedure and found that the inclusion of the second epoch (BD179E1) is responsible for the sub-peak of the $\mu_\alpha$ histogram.
However, since there is no clear evidence of a bad measurement at the second epoch (cf. Figure~\ref{fig:bootstrap_parallax}), we did not take any action such as removing this observation. The availability of pulsar timing proper motions does, however, offer the opportunity to study the effect of applying prior information when conducting the VLBI fitting.

VLBI astrometry is performed in a quasi-inertial reference frame, determined by numerous distant AGNs whose positions (as determined by VLBI observations) are assumed to be fixed.  Examples of realizations of such a reference frame include the International Celestial Reference Frame version 3 (ICRF3\footnote{\url{www.iers.org/IERS/EN/DataProducts/ICRF/ICRF3/icrf3.html}}) and the Radio Fundamental Catalog (RFC\footnote{\url{astrogeo.org/rfc/}}). Timing astrometry, on the other hand, is performed after referencing the pulse ToAs to the barycenter of the solar system, making use of a solar system ephemeris (SSE). Due to the different nature of the reference frames used by VLBI and timing astrometry, a small-scale 3-dimensional rotation between the two types of reference frames is possible, and this rotation could be time-dependent. As a result, small differences in the reference position might be seen between quantities measured using VLBI versus those measured using pulsar timing. Unless the time dependence of the frame misalignment was extremely large, however, the effect on proper motion and (especially) parallax would be extremely small compared to current levels of precision.

The 3-dimensional transformation between ICRF and barycentric frame can be decomposed into a 2-d translation and a 1-d rotation as the displacement is only $\sim1$\,mas level \citep{Wang17}. This can be visualized as the translation and rotation of a local 2-d frame in its surface. The proper motions obtained from VLBI and timing astrometry differ only when there is a noticeable 1-d frame rotation. Under the safe assumption that the angle of the 1-d frame rotation is smaller than 1\,arcmin, the effect of frame transformation on proper motion would be insignificant, and we can in principle make use of the timing proper motions as priors to the VLBI astrometric fitting. The longer time baseline of timing observations promises better precision of proper motion. Therefore if the assumption is met, the application of timing proper motion would potentially improve our parallax estimation.

\begin{figure*}
\centering
\includegraphics[width=14cm]{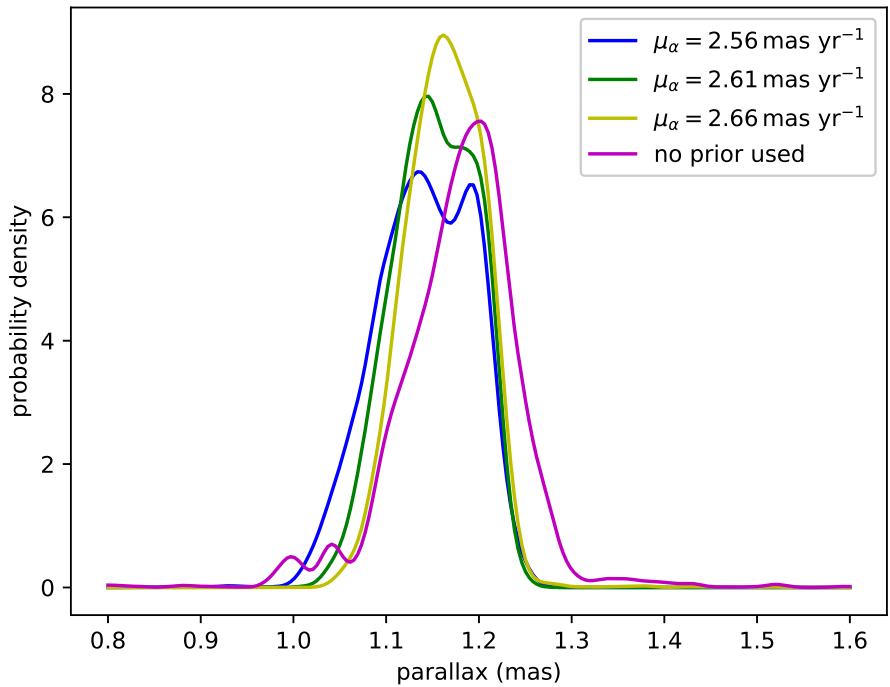}
\caption{Probability density functions of parallax (smoothed-out normalized histogram) out of bootstrapped astrometric fittings using $\mu_\alpha$ priors, in comparison with that of free astrometric fittings.}
\label{fig:timing_prior_pdf}
\end{figure*}

We fixed $\mu_\alpha$ to the 2009, 2016 and 2018 $\mu_\alpha$ respectively (see Table~\ref{tab:astrometric_fits}) and ran bootstrapping again. The resultant parallax probability density functions are shown in Figure~\ref{fig:timing_prior_pdf}. The peak of the normalized histogram of parallax changes slightly with $\mu_\alpha$, as a result of the correlation between parallax and $\mu_\alpha$ (shown by the corresponding error ``ellipse'' in Figure~\ref{fig:pdf_pyramid}). In all cases, the effect of applying the timing proper motion prior is to reduce the most probable parallax value by a small fraction of a standard deviation, and for the most recent timing proper motion results \citep{Desvignes16,Arzoumanian18} the overall parallax uncertainty is reduced.

However, the timing measurements of $\mu_\alpha$ disagree by much more than their formal uncertainties, making it difficult to select the most accurate $\mu_\alpha$ to impose as a prior. Therefore, for simplicity and self-consistency, we use the VLBI results obtained with no priors from the timing proper motion as shown in Table~\ref{tab:astrometric_fits} in following discussion, but note that 1) the application of a well-motivated prior based on timing could further improve the VLBI parallax and hence distance precision, and 2) whichever timing proper motion is chosen, the effect is to reduce the VLBI parallax and hence increase the estimated distance to the pulsar.

\subsection{Galactic path from updated 3D velocity}
\label{subsec:Gal_path}
The orbit of \psr\ through the Galaxy was determined by
\citet{Lazaridis09}, who used their best-fit proper motion
and distance along with a radial velocity of $v_r=44\pm8\,{\rm
  km\,s}^{-1}$ estimated by \citet{Callanan98}. Tracing it back for 10\,Gyr in a model for the Galactic
potential they found that it is only rarely passes close to the Sun; rather
it spends more time out at Galacto-centric distances of 30\,kpc and
oscillating vertically up to 7\,kpc above/below the Galactic plane.  However, \citet{Freire11} used the same data to find different conclusions.  The Galacto-centric distances range from 4\,kpc to 7\,kpc, and the vertical oscillations go to 2\,kpc above/below the Galactic plane.

With our new distance as well as the improved radial velocity
$v_r\, =\, -21.3\, \pm \, 1.6\,{\rm km\,s}^{-1}$ \citep{Mata-Sanchez20}, we can once again repeat this
exercise. We use the \texttt{MWPotential2014} potential from
\texttt{galpy} \citep{Bovy15} for tracing the orbit of the pulsar.
In contrast to \citet{Lazaridis09} and similar to \citet{Freire11}, rather than spending little time near the Sun and orbiting out to 30\,kpc, we find that \psr\ orbits
largely within the Solar circle, going from Galactocentric radii of
3.5--8.5\,kpc on a timescale of $\sim 125\,$Myr.  Similarly, rather
than oscillating to $\pm 7\,$kpc in the vertical direction it only moves
 to $\pm1\,$kpc (the difference between our results and those of \citealt{Freire11} are largely due to the different value of $v_r$ that we assumed).  These results are robust to the choice of
potential: using the same analytic potentials as \citet{Lazaridis09}
gives essentially the same overall orbits.  This is not surprising, as
the 3D space velocity of \psr\ of $105\,{\rm km\,s}^{-1}$ relative to the
Local Standard of Rest (based on the Solar motion of \citealt{Schonrich10})
is largely consistent with the  MSP
distribution \citep{Matthews16}, or maybe a little high, and the
current vertical location in the Galaxy $D\sin b=663\,$pc is likewise
consistent with the MSP distribution \citep{Ng14}, suggesting a
vertical orbit that mimics the overall MSP population with a vertical
scale height of $<1\,$kpc.

\subsection{Constraints on $\dot G$ and dipole gravitational radiation} \label{subsec:alternative_gravity}
Since \citet{Lazaridis09} constrained alternative theories of gravity using timing analysis on \psr, timing parameters including $\dot P_b^{\,\mathrm{obs}}$ (the observed time derivative of orbital period) have been updated by \citet{Desvignes16}. Despite the probable biased parallax presented by the 2016 work (as discussed in Section~\ref{subsec:timing_interplay}), improved measurements for a number of other parameters are expected as a result of longer observation time, most notably (for our purposes) the orbital period derivative and proper motion. The $\dot P_b^{\,\mathrm{obs}}=(6.1\pm0.4)\times 10^{-14}$ reported by \citet{Desvignes16} is consistent with the 2009 counterpart $(5.0\pm1.4)\times 10^{-14}$, but is 3 times more precise. The accuracy of $D_{\dot P_b}$ reported by \citet{Desvignes16} is dominated by the precision of $\dot P_b^{\,\mathrm{obs}}$. 

Besides the improved timing precision, the companion mass $m_c$ and mass ratio $q$ of \psr\ are also better constrained to $0.174\pm0.011$\,\msun\ \citep{Antoniadis16} and $10.44\pm0.11$ \citep{Mata-Sanchez20} respectively.
These, along with our new parallax, allows us to improve the constraints on the $\dot G$ (time derivative of Newton's gravitational constant) and dipole gravitational radiation. The method we adopt here has been developed by \citet{Damour88,Nordtvedt90,Damour91,Lazaridis09,Freire12,Zhu15}.

Contributions to the time variation of orbital period of \psr\ can be summarized as 
\begin{equation}
\label{eq:Pbdot_budget}
{\dot P_b^{\,\mathrm{obs}}} = {\dot P_b^{\,\mathrm{Gal}}} + {\dot P_b^{\,\mathrm{Shk}}} + {\dot P_b^{\,\mathrm{GW}}} + {\dot P_b^{\,{\dot m}}} + {\dot P_b^{\,\mathrm{T}}} + {\dot P_b^{\,\mathrm{ex}}}\,,
\end{equation}
where $\dot P_b^{\,\mathrm{Gal}}$ and $\dot P_b^{\,\mathrm{Shk}}$ are not intrinsic to the \psr\ binary, representing the effect of radial acceleration of \psr\ induced by Galactic gravitational potential \citep{Damour91, Nice95} and transverse motion \citep{Shklovskii70}, respectively; $\dot P_b^{\,\mathrm{GW}}$, $\dot P_b^{\,{\dot m}}$ and $\dot P_b^{\,\mathrm{T}}$ are contributions intrinsic to the binary system resulting from gravitational-wave damping, mass loss of the binary \citep{Damour91,Freire12} and a deformed companion \citep{Smarr76,Freire12}, respectively; $\dot P_b^{\,\mathrm{ex}}$ stands for excess term of non-GR origins. As the non-intrinsic terms of $\dot P_b$ are dependent on distance and proper motion, we are able to refine $\dot P_b^{\,\mathrm{Gal}}$ and $\dot P_b^{\,\mathrm{Shk}}$ with our new astrometric results.

The Shklovskii term can be calculated with 
\begin{equation}
\label{eq:Shk}
\dot P_b^{\,\mathrm{Shk}} = \frac{(\mu_\alpha^2+\mu_\delta^2)D}{c} P_b\,,    
\end{equation}
where $\mu_{\alpha} \equiv \dot{\alpha} \cos\delta$, $D$ is the distance of \psr\ from the Sun and $c$ is the speed of light. 
For the convenience of error propagation, the larger side of uncertainties for parallax and proper motion (Table~\ref{tab:astrometric_fits}) are used as their symmetric uncertainties in the following calculation, i.e. $\varpi=1.21\pm0.08$\,mas, $\mu_{\alpha}=2.67\pm0.09$\,\maspy, $\mu_{\delta}=-25.40\pm0.14$\,\maspy.
Our new parallax and proper motion render $\dot P_b^{\,\mathrm{Shk}} = 68.6\pm4.4$\,\fsps, the uncertainty of which is 63\% of the counterpart in \citet{Lazaridis09}. In the same way as \citet{Zhu15} (and references therein), we updated $\dot P_b^{\,\mathrm{Gal}}=-5.5\pm0.2$\,\fsps\ with our new parallax-based distance to \psr, taking the Sun-GC (Galactic Center) distance to be $R_0=8.122\pm0.031$\,kpc \citep{Gravity-Collaboration18} and circular speed of the local standard of rest to be $\Theta_0=233.3\pm1.4\,{\rm km~s^{-1}}$ \citep{McGaugh18}.
Combining the new mass ratio $q=10.44\pm0.11$ \citep{Mata-Sanchez20} into Equation~(21) from \citet{Lazaridis09}, we acquire $\dot P_b^{\,\mathrm{GW}} = -13\pm1$\,\fsps, where the uncertainty is dominated by the WD mass.

These updated contributions of $\dot P_b$, along with negligible $\dot P_b^{\,{\dot m}}$ and $\dot P_b^{\,\mathrm{T}}$ \citep{Lazaridis09}, give $\dot P_b^{\,\mathrm{ex}} = 10.6\pm6.1$\,\fsps, 2.6 times more precise than the counterpart in \citet{Lazaridis09}. The contributions to $\dot P_b^{\,\mathrm{ex}}$ as well as the derived $\dot P_b^{\,\mathrm{ex}}$ for our work and \citet{Lazaridis09} are summarized in Table~\ref{tab:PbEx_contributions}.

\begin{table*}
	\centering
	\caption{Contributions to the excess orbital decay $\dot P_b^{\,\mathrm{ex}}$}
	\label{tab:PbEx_contributions}
	\begin{tabular}{cccccccc} 
		\hline
		\hline
		& $\dot P_b^{\,\mathrm{obs}}$ & $\dot P_b^{\,\mathrm{Shk}}$ & $\dot P_b^{\,\mathrm{Gal}}$ & $\dot P_b^{\,\mathrm{GW}}$ & $\dot P_b^{\,{\dot m}}$ & $\dot P_b^{\,\mathrm{T}}$ & $\dot P_b^{\,\mathrm{ex}}$ \\
		& \fsps\ & \fsps\ & \fsps\ & \fsps\ & \fsps\ & \fsps\ & \fsps\ \\
		\hline
	This work & 61(4)\tablenotemark{a} & 68.6(4.4) & -5.5(2) & -13(1) & 0 & 0 & 10.6(6.1) \\
	\citet{Lazaridis09} & 50(14) & 70(7) & -5.6(2) & -11(2) & 0 & 0 & -4(16) \\
	\hline
	\end{tabular}
	\tablenotetext{a}{\citet{Desvignes16}}
\end{table*}

As already noted, some alternative theories of gravity demand dipole gravitational radiation and/or a time-dependence to Newton's gravitational constant $G$ in the local universe.
The new $\dot P_b^{\,\mathrm{ex}}$, consistent with zero at 1.7\,$\sigma$ confidence level, does not support alternative theories to GR. 
Nevertheless, we can make use of this measurement to set new limits to dipole gravitational radiation and $\dot G$ with
\begin{equation}
\label{eq:PbdotEx}
\dot P_b^{\,\mathrm{ex}} = \dot P_b^{\,\dot G} + \dot P_b^{\,\mathrm{dp}}\,,
\end{equation}
where $\dot P_b^{\,\dot G}$ and $\dot P_b^{\,\mathrm{dp}}$ represent orbital change caused by $\dot G$ and dipole gravitational radiation, respectively. The relation between $\dot G/G$ and $\dot P_b^{\,\dot G}$ is rewritten from \citet{Damour88,Nordtvedt90,Lazaridis09} as
\begin{equation}
\label{eq:PbdotGdot}
\frac{\dot P_b^{\,\dot G}}{P_b} = -2~\frac{\dot G}{G}\left[1-\left(1+\frac{1}{2}~\frac{1}{q+1}\right)s_p\right] \,,
\end{equation}
where $q=10.44\pm0.11$ \citep{Mata-Sanchez20} is the mass ratio between the pulsar and the companion, $s_p$ stands for the ``sensitivity" of the pulsar depending on its EoS, mass and the theory of gravity in concern \citep{Will93}.
The connection between $\dot P_b^{\,\mathrm{dp}}$ and $\kappa_D$, the putative coupling constant of dipole gravitational radiation, is reproduced from \citet{Lazaridis09} as
\begin{equation}
\label{eq:PbdotDipole}
P_b \dot P_b^{\,\mathrm{dp}} = -4\pi^2 T_\odot m_c~\frac{q}{q+1}~\kappa_D {s_p}^2 \,,
\end{equation}
where $T_\odot=G\msun/c^3=4.9255\,\mu$s, $m_c=0.174\pm0.011\,$\msun\ \citep{Antoniadis16} is the mass of the companion. 
Here we assume the higher-order terms of the ``sensitivities'' of the pulsar and the companion are negligible.

There are two ways to solve $\dot{G}/G$ and $\kappa_D$ from Equations~\ref{eq:PbdotEx}, \ref{eq:PbdotGdot}, \ref{eq:PbdotDipole}. They are both based on the universality of physical laws, i.e. $\dot G/G$ and $\kappa_D$ do not vary in the local universe. The first method is borrowing independent $\dot G/G$ or $\kappa_D$ from other measurements. The second one is using several well timed pulsars to solve or fit (when using more than 2 pulsars) $\dot G/G$ and $\kappa_D$ at the same time, introduced by \citet{Lazaridis09}.

To date, the most stringent limits on $\dot G/G$ are provided by lunar laser ranging (LLR), which yields $\dot G/G=(0.71\pm 1.52)\times 10^{-13}\,{\rm yr^{-1}}$ \citep[95\% confidence level][]{Hofmann18}, and modelling of the orbit of Mercury , which yields $|\dot G|/G=(4\pm 5)\times 10^{-14}\,{\rm yr^{-1}}$ \citep[95\% confidence level][]{Genova18}. The $\dot G/G$ from LLR can be translated into $\dot P_b^{\,\dot G}$, thus assisting us to gauge $\kappa_D$ separately. We use the LLR constraint in preference to that from \citet{Genova18} due to the ambiguity of the sign of the latter. In order to solve $\kappa_D$ in Equation~\ref{eq:PbdotDipole}, we assume $s_p=0.1(m_p/$\msun) (where $m_p=q m_c$), as proposed by \citet{Damour92a} and adopted by \citet{Lazaridis09,Zhu15}. We hence obtain $\dot P_b^{\,\dot G} = -0.19\pm0.20$\,\fsps\ and $\dot P_b^{\,\mathrm{dp}}=10.8\pm6.1$\,\fsps. The latter gives $\kappa_D=(-5.5\pm 6.6)\times 10^{-4}$ (95\% confidence level), which is 3.6 times as precise as the previous effort with \psr\ made utilising the same approach \citep{Lazaridis09}. This estimate of $\kappa_D$ is, however, less precise than the $\kappa_D=(-0.8\pm1.6)\times10^{-4}$ (68\% confidence level) by \citet{Freire12} acquired with the same method while using PSR~J1738$+$0333.

As the second way to solve $\dot{G}/G$ and $\kappa_D$, we combined \psr\ with PSR~J0437$-$4715, PSR~J1738$+$0333 and PSR~J1713$+$0747 to extract $\dot G/G$ and $\kappa_D$, following the method introduced by \citet{Lazaridis09}. The three other pulsars have been used to constrain $\dot G/G$ and $\kappa_D$ \citep{Verbiest08,Deller08,Freire12,Zhu18}. The parameters of the four pulsars we used to derive $\dot G/G$ and $\kappa_D$ are summarized in Table~\ref{tab:4pulsar_parameters}. 
We approach $\dot{G}/G$ and $\kappa_D$ by least-square fitting, and their uncertainties by Monte-Carlo simulation. The marginalized $\dot G/G$ and $\kappa_D$ we obtain are
\begin{equation}
\label{eq:GdotG}
\dot G/G = -1.8^{\,+5.6}_{\,-4.7}\times 10^{-13}\,{\rm yr^{-1}}\,,
\end{equation}
\begin{equation}
\label{eq:kappa_D}
\kappa_D=(-1.7\pm1.7)\times 10^{-4}\,.
\end{equation}
Both values are consistent with zero, and comparably precise to (and slightly more conservative than) $\kappa_D=(-0.7\pm2.2)\times10^{-4}$ and $\dot{G}/G=(-1\pm9)\times10^{-13}\,{\rm yr^{-1}}$ measured at 95\% confidence by \citet{Zhu18}.
We note that we have adopted the relation $s_p=0.1(m_p/$\msun), and hence these two estimates are dependent on this assumed $s_p$ relation.
The mathematical formalism of $s_p$ hinges further on the EoS of NSs, which will be better constrained by NICER \citep{Bogdanov19,Bogdanov19a} and the gravitational-wave observatories \citep{Annala18} in the years to come.

\begin{table*}
	\centering
	\caption{Parameters of MSPs for estimating $\dot G/G$ and $\kappa_D$} 
	\label{tab:4pulsar_parameters}
	\begin{tabular}{ccccccc} 
		\hline
		\hline
	pulsar & $P_b$ & $\dot P_b^{\,\mathrm{ex}}$ & $m_p$ & $m_c$ & $q$ & references  \\
		& d & \fsps\ & \msun\ & \msun\ &  & \\
		\hline
	PSR~J0437$-$4715 & 5.74 & 12(32)\tablenotemark{a} & 1.44(7) & 0.224(7) & $-$ & \citet{Reardon16,Deller08} \\
	\psr\ (this work) & 0.60 & 10.6(6.1) & $-$ & 0.174(11) & 10.44(11) & references in this paper \\
    PSR~J1713$+$0747 & 67.83 & 30(150) & 1.33(10) & 0.290(11) & $-$ & \citet{Zhu18} \\
    PSR~J1738$+$0333 & 0.35 & 2.0(3.7) & 1.46(6) & $-$ & 8.1(2) & \citet{Freire12} \\
	\hline
	\end{tabular}
	\tablenotetext{a}{We derived $\dot P_b^{\,\mathrm{ex}}$ for PSR~J0437$-$4715 with the results from \citet{Reardon16,Deller08}.}
\end{table*}

\section{Conclusions and future prospects}
\label{sec:conclusions}
\begin{enumerate}
\item This paper reports new VLBI astrometry of \psr\ (Table~\ref{tab:astrometric_fits}). Our new distance to \psr, $0.83^{+0.06}_{-0.02}$\,kpc, is the most precise to date and consistent with major measurements. We present the first VLBI-based absolute position for \psr, which paves the road for the frame link between the quasi-static International Celestial Reference Frame used by VLBI and the solar-system frame used by pulsar timing.

\item Using our new distance and proper motion, we reduce the uncertainty of the Shklovskii term in Equation~\ref{eq:Pbdot_budget}. On top of that, we set new constraints on the fractional time derivative of the Newton's gravitational constant $\dot G/G$ for the local universe and the coupling constant for dipolar gravitational radiation $\kappa_D$, combining three other millisecond pulsars, PSR~J0437$-$4715, PSR~J1738$+$0333 and PSR~J1713$+$0747. The new $\kappa_D$ is comparable to the most stringent constraint.

\item As is shown in Table~\ref{tab:4pulsar_parameters}, among the four pulsars, the $\dot{P_b}^{\mathrm{ex}}$ of \psr\ stands out with $>1\,\sigma$ offset from zero, which effectively brings the best-fit $\kappa_D$ away from zero. If we only use the other three pulsars and re-do the analysis, we obtain $\kappa_D=(-0.8^{\,+1.9}_{\,-1.7})\times10^{-4}$, where the uncertainty increases but the best-fit $\kappa_D$ becomes more consistent with zero. Therefore, whether $\dot{P_b}^{\mathrm{ex}}$ of \psr\ will converge to zero with future timing and VLBI observations of \psr\ is essential for the $\kappa_D$ test using pulsar-WD binaries. 
Assuming GR is correct, the 1.7\,$\sigma$ offset of $\dot P_b^{\,\mathrm{ex}}$ from zero implies
that the value of either our VLBI parallax and/or $\dot{P_b}^{\mathrm{obs}}$ is too high (given that the other contributing terms of $\dot{P_b}$ vary marginally).
This mild tension will be re-visited with a better $\dot{P_b}^{\mathrm{obs}}$ based on longer timing observations. 
As noted in Section~\ref{subsec:timing_interplay}, we see that applying constraints to the proper motion of \psr\ based on pulsar timing act to reduce the VLBI parallax, albeit within the current uncertainties, which would already mitigate the tension somewhat.
It is likely that the timing proper motions of \psr\ acquired independently with EPTA and NANOGrav will converge to a value with negligible uncertainty in the future, allowing us to confidently apply this as a prior to our VLBI fit and improve our parallax estimate for \psr. 
Furthermore, when the uncertainties of both $\dot P_b^{\,\mathrm{obs}}$ and the distance to \psr\ are improved by a factor of 4 with new timing and VLBI observations, we are able to estimate the next uncertainty contributor to $\dot P_b^{\,\mathrm{ex}}$ - the WD mass (as well as the pulsar mass), assuming $\dot P_b^{\,\mathrm{ex}}=0$. This independent WD mass will help refine the relation between WD mass and WD atmospheric parameters in the helium-WD regime.

\item Looking into the future, the uncertainty of $\dot P_b^{\,\mathrm{obs}}$ for \psr\ will quickly vanish \citep[as $t^{\,-2.5}$,][]{Bell96}, reducing the uncertainty of $\dot P_b^{\,\mathrm{ex}}$ to $\approx4.6$\,\fsps\ in $\approx10$ years. At that time, the Shklovskii term would become the leading error source of $\dot P_b^{\,\mathrm{ex}}$ for \psr.
Inside the Shklovskii term, the distance (or parallax) dominates the error budget as the uncertainty of parallax improves as $t^{\,-0.5}$. That means distance uncertainty will eventually become the biggest barrier against better constraints on alternative theories of gravity. This will be the same for the analysis of most\footnote{For some systems, especially PSR~J1738$+$0333, the relatively low precision on the mass measurements - not the uncertainty on the distance - will be the main barrier towards improvement of the test.} other MSPs.
If we reduce the uncertainty of $\dot P_b^{\,\mathrm{obs}}$ to zero for each of the four above-mentioned pulsars (\psr, PSR~J0437$-$4715, PSR~J1738$+$0333 and PSR~J1713$+$0747) and re-derive $\dot{G}/G$ and $\kappa_D$, we find the $1\,\sigma$ uncertainty reduces to $\leq1.5\times10^{-13}\,{\rm yr^{-1}}$ for $\dot{G}/G$ and $\leq1.0\times10^{-4}$ for $\kappa_D$. This simulation shows significantly better constraints on $\dot{G}/G$ and $\kappa_D$ can be made with continuous efforts on pulsar timing for $\approx10$ years. Beyond that, in order to further this study, we need to focus on improving the precision of distances to the pulsars of use.
\end{enumerate}

\acknowledgments
HD is supported by the ACAMAR (Australia-ChinA ConsortiuM for Astrophysical Research) scholarship, which is partly funded by the China Scholarship Council (CSC). DLK and TJWL were supported by the NANOGrav Physics Frontiers Center, which is supported by the National Science Foundation award 1430284. Part of this research was carried out at the Jet Propulsion Laboratory, California Institute of Technology, under a contract with the National Aeronautics and Space Administration. 
The authors thank Norbert Wex and the anonymous referee for their important comments on this paper.
The data of this work come from the Very Long Baseline Array (VLBA), which is operated by the National Radio Astronomy Observatory (NRAO). The NRAO is a facility of the National Science Foundation operated under cooperative agreement by Associated Universities, Inc.
This work made use of the Swinburne University of Technology software correlator, developed as part of the Australian Major National Research Facilities Programme and operated under license.

\software{DiFX \citep{Deller11a}, ParselTongue \citep{Kettenis06},
AIPS \citep{Greisen03},
DIFMAP \citep{Shepherd94},
psrvlbireduce (\url{https://github.com/dingswin/psrvlbireduce}),
pmpar (\url{https://github.com/walterfb/pmpar})}

\newpage
\bibliographystyle{aasjournal}
\bibliography{J1012,haoding}

\begin{thebibliography}{}
\expandafter\ifx\csname natexlab\endcsname\relax\def\natexlab#1{#1}\fi
\providecommand{\url}[1]{\href{#1}{#1}}

\bibitem[{{Alpar} {et~al.}(1982){Alpar}, {Cheng}, {Ruderman}, \&
  {Shaham}}]{Alpar82}
{Alpar}, M.~A., {Cheng}, A.~F., {Ruderman}, M.~A., \& {Shaham}, J. 1982, \nat,
  300, 728

\bibitem[{Annala {et~al.}(2018)Annala, Gorda, Kurkela, \& Vuorinen}]{Annala18}
Annala, E., Gorda, T., Kurkela, A., \& Vuorinen, A. 2018, Physical review
  letters, 120, 172703

\bibitem[{{Antoniadis} {et~al.}(2016){Antoniadis}, {Tauris}, {Ozel}, {Barr},
  {Champion}, \& {Freire}}]{Antoniadis16}
{Antoniadis}, J., {Tauris}, T.~M., {Ozel}, F., {et~al.} 2016, arXiv e-prints,
  arXiv:1605.01665

\bibitem[{{Antoniadis} {et~al.}(2013){Antoniadis}, {Freire}, {Wex}, {Tauris},
  {Lynch}, {van Kerkwijk}, {Kramer}, {Bassa}, {Dhillon}, {Driebe}, {Hessels},
  {Kaspi}, {Kondratiev}, {Langer}, {Marsh}, {McLaughlin}, {Pennucci}, {Ransom},
  {Stairs}, {van Leeuwen}, {Verbiest}, \& {Whelan}}]{Antoniadis13}
{Antoniadis}, J., {Freire}, P.~C.~C., {Wex}, N., {et~al.} 2013, Science, 340,
  448

\bibitem[{Archibald {et~al.}(2018)Archibald, Gusinskaia, Hessels, Deller,
  Kaplan, Lorimer, Lynch, Ransom, \& Stairs}]{Archibald18}
Archibald, A.~M., Gusinskaia, N.~V., Hessels, J.~W., {et~al.} 2018, Nature,
  559, 73

\bibitem[{{Arzoumanian} {et~al.}(2018){Arzoumanian}, {Baker}, {Brazier},
  {Burke-Spolaor}, {Chamberlin}, {Chatterjee}, {Christy}, {Cordes}, {Cornish},
  {Crawford}, {Thankful Cromartie}, {Crowter}, {DeCesar}, {Demorest}, {Dolch},
  {Ellis}, {Ferdman}, {Ferrara}, {Folkner}, {Fonseca}, {Garver-Daniels},
  {Gentile}, {Haas}, {Hazboun}, {Huerta}, {Islo}, {Jones}, {Jones}, {Kaplan},
  {Kaspi}, {Lam}, {Lazio}, {Levin}, {Lommen}, {Lorimer}, {Luo}, {Lynch},
  {Madison}, {McLaughlin}, {McWilliams}, {Mingarelli}, {Ng}, {Nice}, {Park},
  {Pennucci}, {Pol}, {Ransom}, {Ray}, {Rasskazov}, {Siemens}, {Simon},
  {Spiewak}, {Stairs}, {Stinebring}, {Stovall}, {Swiggum}, {Taylor},
  {Vallisneri}, {van Haasteren}, {Vigeland}, {Zhu}, \& {NANOGrav
  Collaboration}}]{Arzoumanian18}
{Arzoumanian}, Z., {Baker}, P.~T., {Brazier}, A., {et~al.} 2018, \apj, 859, 47

\bibitem[{{Bartel} {et~al.}(1986){Bartel}, {Herring}, {Ratner}, {Shapiro}, \&
  {Corey}}]{Bartel86}
{Bartel}, N., {Herring}, T.~A., {Ratner}, M.~I., {Shapiro}, I.~I., \& {Corey},
  B.~E. 1986, \nat, 319, 733

\bibitem[{{Bell} \& {Bailes}(1996)}]{Bell96}
{Bell}, J.~F., \& {Bailes}, M. 1996, \apjl, 456, L33

\bibitem[{{Bhat} {et~al.}(2004){Bhat}, {Cordes}, {Camilo}, {Nice}, \&
  {Lorimer}}]{Bhat04}
{Bhat}, N.~D.~R., {Cordes}, J.~M., {Camilo}, F., {Nice}, D.~J., \& {Lorimer},
  D.~R. 2004, \apj, 605, 759

\bibitem[{Bogdanov {et~al.}(2019{\natexlab{a}})Bogdanov, Guillot, Ray, Wolff,
  Chakrabarty, Ho, Kerr, Lamb, Lommen, Ludlam, {et~al.}}]{Bogdanov19}
Bogdanov, S., Guillot, S., Ray, P.~S., {et~al.} 2019{\natexlab{a}}, The
  Astrophysical Journal Letters, 887, L25

\bibitem[{Bogdanov {et~al.}(2019{\natexlab{b}})Bogdanov, Lamb, Mahmoodifar,
  Miller, Morsink, Riley, Strohmayer, Tung, Watts, Dittmann,
  {et~al.}}]{Bogdanov19a}
Bogdanov, S., Lamb, F.~K., Mahmoodifar, S., {et~al.} 2019{\natexlab{b}}, The
  Astrophysical Journal Letters, 887, L26

\bibitem[{Bovy(2015)}]{Bovy15}
Bovy, J. 2015, \apjs, 216, 29

\bibitem[{{Burgay} {et~al.}(2003){Burgay}, {D'Amico}, {Possenti}, {Manchester},
  {Lyne}, {Joshi}, {McLaughlin}, {Kramer}, {Sarkissian}, {Camilo}, {Kalogera},
  {Kim}, \& {Lorimer}}]{Burgay03}
{Burgay}, M., {D'Amico}, N., {Possenti}, A., {et~al.} 2003, \nat, 426, 531

\bibitem[{{Callanan} {et~al.}(1998){Callanan}, {Garnavich}, \&
  {Koester}}]{Callanan98}
{Callanan}, P.~J., {Garnavich}, P.~M., \& {Koester}, D. 1998, \mnras, 298, 207

\bibitem[{{Chatterjee} {et~al.}(2009){Chatterjee}, {Brisken}, {Vlemmings},
  {Goss}, {Lazio}, {Cordes}, {Thorsett}, {Fomalont}, {Lyne}, \&
  {Kramer}}]{Chatterjee09}
{Chatterjee}, S., {Brisken}, W.~F., {Vlemmings}, W.~H.~T., {et~al.} 2009, \apj,
  698, 250

\bibitem[{Damour \& Esposito-Farese(1992)}]{Damour92a}
Damour, T., \& Esposito-Farese, G. 1992, Classical and Quantum Gravity, 9, 2093

\bibitem[{{Damour} {et~al.}(1988){Damour}, {Gibbons}, \& {Taylor}}]{Damour88}
{Damour}, T., {Gibbons}, G.~W., \& {Taylor}, J.~H. 1988, Physical Review
  Letters, 61, 1151

\bibitem[{{Damour} \& {Taylor}(1991)}]{Damour91}
{Damour}, T., \& {Taylor}, J.~H. 1991, \apj, 366, 501

\bibitem[{{Damour} \& {Taylor}(1992)}]{Damour92}
---. 1992, \prd, 45, 1840

\bibitem[{{Deller} {et~al.}(2008){Deller}, {Verbiest}, {Tingay}, \&
  {Bailes}}]{Deller08}
{Deller}, A.~T., {Verbiest}, J.~P.~W., {Tingay}, S.~J., \& {Bailes}, M. 2008,
  \apjl, 685, L67

\bibitem[{{Deller} {et~al.}(2011{\natexlab{a}}){Deller}, {Brisken},
  {Chatterjee}, {Cordes}, {Goss}, {Janssen}, {Kovalev}, {Lazio}, {Petrov}, \&
  {Stappers}}]{Deller11}
{Deller}, A.~T., {Brisken}, W.~F., {Chatterjee}, S., {et~al.}
  2011{\natexlab{a}}, in 20th Meeting of the European VLBI Group for Geodesy
  and Astronomy, held in Bonn, Germany, March 29-30, 2011, Eds: W. Alef, S.
  Bernhart, and A. Nothnagel, Institut f{\"u}r Geod{\"a}sie und Geoinformation,
  Rheinischen Friedrich-Wilhelms-Universit{\"a}t Bonn, p. 178-182, ed.
  W.~{Alef}, S.~{Bernhart}, \& A.~{Nothnagel}, 178--182

\bibitem[{{Deller} {et~al.}(2011{\natexlab{b}}){Deller}, {Brisken}, {Phillips},
  {Morgan}, {Alef}, {Cappallo}, {Middelberg}, {Romney}, {Rottmann}, {Tingay},
  \& {Wayth}}]{Deller11a}
{Deller}, A.~T., {Brisken}, W.~F., {Phillips}, C.~J., {et~al.}
  2011{\natexlab{b}}, \pasp, 123, 275

\bibitem[{{Deller} {et~al.}(2016){Deller}, {Vigeland}, {Kaplan}, {Goss},
  {Brisken}, {Chatterjee}, {Cordes}, {Janssen}, {Lazio}, {Petrov}, {Stappers},
  \& {Lyne}}]{Deller16}
{Deller}, A.~T., {Vigeland}, S.~J., {Kaplan}, D.~L., {et~al.} 2016, \apj, 828,
  8

\bibitem[{{Deller} {et~al.}(2019){Deller}, {Goss}, {Brisken}, {Chatterjee},
  {Cordes}, {Janssen}, {Kovalev}, {Lazio}, {Petrov}, {Stappers}, \&
  {Lyne}}]{Deller19}
{Deller}, A.~T., {Goss}, W.~M., {Brisken}, W.~F., {et~al.} 2019, \apj, 875, 100

\bibitem[{{Desvignes} {et~al.}(2016){Desvignes}, {Caballero}, {Lentati},
  {Verbiest}, {Champion}, {Stappers}, {Janssen}, {Lazarus}, {Os{\l}owski},
  {Babak}, {Bassa}, {Brem}, {Burgay}, {Cognard}, {Gair}, {Graikou},
  {Guillemot}, {Hessels}, {Jessner}, {Jordan}, {Karuppusamy}, {Kramer},
  {Lassus}, {Lazaridis}, {Lee}, {Liu}, {Lyne}, {McKee}, {Mingarelli},
  {Perrodin}, {Petiteau}, {Possenti}, {Purver}, {Rosado}, {Sanidas}, {Sesana},
  {Shaifullah}, {Smits}, {Taylor}, {Theureau}, {Tiburzi}, {van Haasteren}, \&
  {Vecchio}}]{Desvignes16}
{Desvignes}, G., {Caballero}, R.~N., {Lentati}, L., {et~al.} 2016, \mnras, 458,
  3341

\bibitem[{{Detweiler}(1979)}]{Detweiler79}
{Detweiler}, S. 1979, \apj, 234, 1100

\bibitem[{{Driebe} {et~al.}(1998){Driebe}, {Schoenberner}, {Bloecker}, \&
  {Herwig}}]{Driebe98}
{Driebe}, T., {Schoenberner}, D., {Bloecker}, T., \& {Herwig}, F. 1998, \aap,
  339, 123

\bibitem[{Freire {et~al.}(2011)Freire, Bassa, Wex, Stairs, Champion, Ransom,
  Lazarus, Kaspi, Hessels, Kramer, {et~al.}}]{Freire11}
Freire, P., Bassa, C., Wex, N., {et~al.} 2011, Monthly Notices of the Royal
  Astronomical Society, 412, 2763

\bibitem[{{Freire} {et~al.}(2012){Freire}, {Wex}, {Esposito-Far{\`e}se},
  {Verbiest}, {Bailes}, {Jacoby}, {Kramer}, {Stairs}, {Antoniadis}, \&
  {Janssen}}]{Freire12}
{Freire}, P.~C.~C., {Wex}, N., {Esposito-Far{\`e}se}, G., {et~al.} 2012,
  \mnras, 423, 3328

\bibitem[{{Gaia Collaboration} {et~al.}(2016){Gaia Collaboration}, {Prusti},
  {de Bruijne}, {Brown}, {Vallenari}, {Babusiaux}, {Bailer-Jones}, {Bastian},
  {Biermann}, {Evans}, \& et~al.}]{Gaia-Collaboration16}
{Gaia Collaboration}, {Prusti}, T., {de Bruijne}, J.~H.~J., {et~al.} 2016,
  \aap, 595, A1

\bibitem[{{Gaia Collaboration} {et~al.}(2018){Gaia Collaboration}, {Mignard},
  {Klioner}, {Lindegren}, {Hern{\'a}ndez}, {Bastian}, {Bombrun}, {Hobbs},
  {Lammers}, {Michalik}, \& et~al.}]{Gaia-Collaboration18}
{Gaia Collaboration}, {Mignard}, F., {Klioner}, S.~A., {et~al.} 2018, \aap,
  616, A14

\bibitem[{Genova {et~al.}(2018)Genova, Mazarico, Goossens, Lemoine, Neumann,
  Smith, \& Zuber}]{Genova18}
Genova, A., Mazarico, E., Goossens, S., {et~al.} 2018, Nature communications,
  9, 289

\bibitem[{{Gravity Collaboration} {et~al.}(2018){Gravity Collaboration},
  {Abuter}, {Amorim}, {Anugu}, {Baub{\"o}ck}, {Benisty}, {Berger}, {Blind},
  {Bonnet}, {Brandner}, {Buron}, {Collin}, {Chapron}, {Cl{\'e}net}, {Coud{\'e}
  Du Foresto}, {de Zeeuw}, {Deen}, {Delplancke-Str{\"o}bele}, {Dembet},
  {Dexter}, {Duvert}, {Eckart}, {Eisenhauer}, {Finger}, {F{\"o}rster
  Schreiber}, {F{\'e}dou}, {Garcia}, {Garcia Lopez}, {Gao}, {Gendron},
  {Genzel}, {Gillessen}, {Gordo}, {Habibi}, {Haubois}, {Haug}, {Hau{\ss}mann},
  {Henning}, {Hippler}, {Horrobin}, {Hubert}, {Hubin}, {Jimenez Rosales},
  {Jochum}, {Jocou}, {Kaufer}, {Kellner}, {Kendrew}, {Kervella}, {Kok},
  {Kulas}, {Lacour}, {Lapeyr{\`e}re}, {Lazareff}, {Le Bouquin}, {L{\'e}na},
  {Lippa}, {Lenzen}, {M{\'e}rand}, {M{\"u}ler}, {Neumann}, {Ott}, {Palanca},
  {Paumard}, {Pasquini}, {Perraut}, {Perrin}, {Pfuhl}, {Plewa}, {Rabien},
  {Ram{\'{\i}}rez}, {Ramos}, {Rau}, {Rodr{\'{\i}}guez-Coira}, {Rohloff},
  {Rousset}, {Sanchez-Bermudez}, {Scheithauer}, {Sch{\"o}ller}, {Schuler},
  {Spyromilio}, {Straub}, {Straubmeier}, {Sturm}, {Tacconi}, {Tristram},
  {Vincent}, {von Fellenberg}, {Wank}, {Waisberg}, {Widmann}, {Wieprecht},
  {Wiest}, {Wiezorrek}, {Woillez}, {Yazici}, {Ziegler}, \&
  {Zins}}]{Gravity-Collaboration18}
{Gravity Collaboration}, {Abuter}, R., {Amorim}, A., {et~al.} 2018, \aap, 615,
  L15

\bibitem[{{Greisen}(2003)}]{Greisen03}
{Greisen}, E.~W. 2003, in Astrophysics and Space Science Library, Vol. 285,
  Information Handling in Astronomy - Historical Vistas, ed. A.~{Heck}, 109

\bibitem[{{Hofmann} \& {M{\"u}ller}(2018)}]{Hofmann18}
{Hofmann}, F., \& {M{\"u}ller}, J. 2018, Classical and Quantum Gravity, 35,
  035015

\bibitem[{{Jennings} {et~al.}(2018){Jennings}, {Kaplan}, {Chatterjee},
  {Cordes}, \& {Deller}}]{Jennings18}
{Jennings}, R.~J., {Kaplan}, D.~L., {Chatterjee}, S., {Cordes}, J.~M., \&
  {Deller}, A.~T. 2018, \apj, 864, 26

\bibitem[{{Kettenis} {et~al.}(2006){Kettenis}, {van Langevelde}, {Reynolds}, \&
  {Cotton}}]{Kettenis06}
{Kettenis}, M., {van Langevelde}, H.~J., {Reynolds}, C., \& {Cotton}, B. 2006,
  in Astronomical Society of the Pacific Conference Series, Vol. 351,
  Astronomical Data Analysis Software and Systems XV, ed. C.~{Gabriel},
  C.~{Arviset}, D.~{Ponz}, \& S.~{Enrique}, 497

\bibitem[{{Lange} {et~al.}(2001){Lange}, {Camilo}, {Wex}, {Kramer}, {Backer},
  {Lyne}, \& {Doroshenko}}]{Lange01}
{Lange}, C., {Camilo}, F., {Wex}, N., {et~al.} 2001, \mnras, 326, 274

\bibitem[{{Lazaridis} {et~al.}(2009){Lazaridis}, {Wex}, {Jessner}, {Kramer},
  {Stappers}, {Janssen}, {Desvignes}, {Purver}, {Cognard}, {Theureau}, {Lyne},
  {Jordan}, \& {Zensus}}]{Lazaridis09}
{Lazaridis}, K., {Wex}, N., {Jessner}, A., {et~al.} 2009, \mnras, 400, 805

\bibitem[{{Lobanov}(1998)}]{Lobanov98}
{Lobanov}, A.~P. 1998, \aap, 330, 79

\bibitem[{{Lorimer} \& {Kramer}(2012)}]{Lorimer12}
{Lorimer}, D.~R., \& {Kramer}, M. 2012, {Handbook of Pulsar Astronomy}
  (Cambridge University Press)

\bibitem[{{Lutz} \& {Kelker}(1973)}]{Lutz73}
{Lutz}, T.~E., \& {Kelker}, D.~H. 1973, \pasp, 85, 573

\bibitem[{{Lyne} \& {Rickett}(1968)}]{Lyne68}
{Lyne}, A.~G., \& {Rickett}, B.~J. 1968, \nat, 219, 1339

\bibitem[{{Lyutikov} \& {Thompson}(2005)}]{Lyutikov05}
{Lyutikov}, M., \& {Thompson}, C. 2005, \apj, 634, 1223

\bibitem[{{Madison} {et~al.}(2013){Madison}, {Chatterjee}, \&
  {Cordes}}]{Madison13}
{Madison}, D.~R., {Chatterjee}, S., \& {Cordes}, J.~M. 2013, \apj, 777, 104

\bibitem[{Mata~S{\'a}nchez {et~al.}(2020)Mata~S{\'a}nchez, Istrate, van
  Kerkwijk, Breton, \& Kaplan}]{Mata-Sanchez20}
Mata~S{\'a}nchez, D., Istrate, A., van Kerkwijk, M., Breton, R., \& Kaplan, D.
  2020, arXiv preprint arXiv:2004.02901

\bibitem[{Matthews {et~al.}(2016)Matthews, Nice, Fonseca, Arzoumanian, Crowter,
  Demorest, Dolch, Ellis, Ferdman, Gonzalez, {et~al.}}]{Matthews16}
Matthews, A.~M., Nice, D.~J., Fonseca, E., {et~al.} 2016, \apj, 818, 92

\bibitem[{{McGaugh}(2018)}]{McGaugh18}
{McGaugh}, S.~S. 2018, Research Notes of the American Astronomical Society, 2,
  156

\bibitem[{Ng {et~al.}(2014)Ng, Bailes, Bates, Bhat, Burgay, Burke-Spolaor,
  Champion, Coster, Johnston, Keith, {et~al.}}]{Ng14}
Ng, C., Bailes, M., Bates, S., {et~al.} 2014, \mnras, 439, 1865

\bibitem[{{Nicastro} {et~al.}(1995){Nicastro}, {Lyne}, {Lorimer}, {Harrison},
  {Bailes}, \& {Skidmore}}]{Nicastro95}
{Nicastro}, L., {Lyne}, A.~G., {Lorimer}, D.~R., {et~al.} 1995, \mnras, 273,
  L68

\bibitem[{{Nice} \& {Taylor}(1995)}]{Nice95}
{Nice}, D.~J., \& {Taylor}, J.~H. 1995, \apj, 441, 429

\bibitem[{{Nordtvedt}(1990)}]{Nordtvedt90}
{Nordtvedt}, K. 1990, Physical Review Letters, 65, 953

\bibitem[{Perera {et~al.}(2019)Perera, DeCesar, Demorest, Kerr, Lentati, Nice,
  Os{\l}owski, Ransom, Keith, Arzoumanian, {et~al.}}]{Perera19}
Perera, B., DeCesar, M., Demorest, P., {et~al.} 2019, Monthly Notices of the
  Royal Astronomical Society, 490, 4666

\bibitem[{{Petrov} {et~al.}(2008){Petrov}, {Kovalev}, {Fomalont}, \&
  {Gordon}}]{Petrov08}
{Petrov}, L., {Kovalev}, Y.~Y., {Fomalont}, E.~B., \& {Gordon}, D. 2008, \aj,
  136, 580

\bibitem[{Plavin {et~al.}(2019)Plavin, Kovalev, Pushkarev, \&
  Lobanov}]{Plavin19}
Plavin, A., Kovalev, Y., Pushkarev, A., \& Lobanov, A. 2019, Monthly Notices of
  the Royal Astronomical Society, 485, 1822

\bibitem[{Reardon {et~al.}(2016)Reardon, Hobbs, Coles, Levin, Keith, Bailes,
  Bhat, Burke-Spolaor, Dai, Kerr, {et~al.}}]{Reardon16}
Reardon, D., Hobbs, G., Coles, W., {et~al.} 2016, Monthly Notices of the Royal
  Astronomical Society, 455, 1751

\bibitem[{Sch{\"o}nrich {et~al.}(2010)Sch{\"o}nrich, Binney, \&
  Dehnen}]{Schonrich10}
Sch{\"o}nrich, R., Binney, J., \& Dehnen, W. 2010, \mnras, 403, 1829

\bibitem[{Shannon \& Cordes(2010)}]{Shannon10}
Shannon, R.~M., \& Cordes, J.~M. 2010, The Astrophysical Journal, 725, 1607

\bibitem[{{Shepherd} {et~al.}(1994){Shepherd}, {Pearson}, \&
  {Taylor}}]{Shepherd94}
{Shepherd}, M.~C., {Pearson}, T.~J., \& {Taylor}, G.~B. 1994, in \baas,
  Vol.~26, Bulletin of the American Astronomical Society, 987--989

\bibitem[{{Shklovskii}(1970)}]{Shklovskii70}
{Shklovskii}, I.~S. 1970, \sovast, 13, 562

\bibitem[{{Smarr} \& {Blandford}(1976)}]{Smarr76}
{Smarr}, L.~L., \& {Blandford}, R. 1976, \apj, 207, 574

\bibitem[{{Sokolovsky} {et~al.}(2011){Sokolovsky}, {Kovalev}, {Pushkarev}, \&
  {Lobanov}}]{Sokolovsky11}
{Sokolovsky}, K.~V., {Kovalev}, Y.~Y., {Pushkarev}, A.~B., \& {Lobanov}, A.~P.
  2011, \aap, 532, A38

\bibitem[{{van Kerkwijk} {et~al.}(2004){van Kerkwijk}, Bassa, Jacoby, \&
  Jonker}]{van-Kerkwijk04}
{van Kerkwijk}, M.~H., Bassa, C., Jacoby, B., \& Jonker, P. 2004, arXiv
  preprint astro-ph/0405283

\bibitem[{{van Kerkwijk} {et~al.}(1996){van Kerkwijk}, {Bergeron}, \&
  {Kulkarni}}]{van-Kerkwijk96}
{van Kerkwijk}, M.~H., {Bergeron}, P., \& {Kulkarni}, S.~R. 1996, \apjl, 467,
  L89

\bibitem[{Verbiest {et~al.}(2016)Verbiest, Lentati, Hobbs, van Haasteren,
  Demorest, Janssen, Wang, Desvignes, Caballero, Keith, {et~al.}}]{Verbiest16}
Verbiest, J., Lentati, L., Hobbs, G., {et~al.} 2016, Monthly Notices of the
  Royal Astronomical Society, 458, 1267

\bibitem[{Verbiest {et~al.}(2008)Verbiest, Bailes, van Straten, Hobbs, Edwards,
  Manchester, Bhat, Sarkissian, Jacoby, \& Kulkarni}]{Verbiest08}
Verbiest, J.~P., Bailes, M., van Straten, W., {et~al.} 2008, The Astrophysical
  Journal, 679, 675

\bibitem[{{Vigeland} {et~al.}(2018){Vigeland}, {Deller}, {Kaplan}, {Istrate},
  {Stappers}, \& {Tauris}}]{Vigeland18}
{Vigeland}, S.~J., {Deller}, A.~T., {Kaplan}, D.~L., {et~al.} 2018, \apj, 855,
  122

\bibitem[{{Wang} {et~al.}(2017){Wang}, {Coles}, {Hobbs}, {Shannon},
  {Manchester}, {Kerr}, {Yuan}, {Wang}, {Bailes}, {Bhat}, {Dai}, {Dempsey},
  {Keith}, {Lasky}, {Levin}, {Os{\l}owski}, {Ravi}, {Reardon}, {Rosado},
  {Russell}, {Spiewak}, {van Straten}, {Toomey}, {Wen}, {You}, \&
  {Zhu}}]{Wang17}
{Wang}, J.~B., {Coles}, W.~A., {Hobbs}, G., {et~al.} 2017, \mnras, 469, 425

\bibitem[{{Will}(1993)}]{Will93}
{Will}, C.~M. 1993, {Theory and Experiment in Gravitational Physics} (Cambridge
  University Press), 396

\bibitem[{{Yao} {et~al.}(2017){Yao}, {Manchester}, \& {Wang}}]{Yao17}
{Yao}, J.~M., {Manchester}, R.~N., \& {Wang}, N. 2017, \apj, 835, 29

\bibitem[{Zhu {et~al.}(2018)Zhu, Desvignes, Wex, Caballero, Champion, Demorest,
  Ellis, Janssen, Kramer, Krieger, {et~al.}}]{Zhu18}
Zhu, W., Desvignes, G., Wex, N., {et~al.} 2018, Monthly Notices of the Royal
  Astronomical Society, 482, 3249

\bibitem[{{Zhu} {et~al.}(2015){Zhu}, {Stairs}, {Demorest}, {Nice}, {Ellis},
  {Ransom}, {Arzoumanian}, {Crowter}, {Dolch}, {Ferdman}, {Fonseca},
  {Gonzalez}, {Jones}, {Jones}, {Lam}, {Levin}, {McLaughlin}, {Pennucci},
  {Stovall}, \& {Swiggum}}]{Zhu15}
{Zhu}, W.~W., {Stairs}, I.~H., {Demorest}, P.~B., {et~al.} 2015, \apj, 809, 41

\end{thebibliography}



\end{document}